Taylor & Francis
Taylor & Francis Group

PERSPECTIVE

# Integrative biological simulation praxis: Considerations from physics, philosophy, and data/model curation practices

Gopal P. Sarma [ID][a] and Victor Faundez [ID][a,b]

[a]School of Medicine, Emory University, Atlanta, GA, USA; [b]Department of Cell Biology, Emory University, Atlanta, GA, USA

**ABSTRACT**

Integrative biological simulations have a varied and controversial history in the biological sciences. From computational models of organelles, cells, and simple organisms, to physiological models of tissues, organ systems, and ecosystems, a diverse array of biological systems have been the target of large-scale computational modeling efforts. Nonetheless, these research agendas have yet to prove decisively their value among the broader community of theoretical and experimental biologists. In this commentary, we examine a range of philosophical and practical issues relevant to understanding the potential of integrative simulations. We discuss the role of theory and modeling in different areas of physics and suggest that certain sub-disciplines of physics provide useful cultural analogies for imagining the future role of simulations in biological research. We examine philosophical issues related to modeling which consistently arise in discussions about integrative simulations and suggest a pragmatic viewpoint that balances a belief in philosophy with the recognition of the relative infancy of our state of philosophical understanding. Finally, we discuss community workflow and publication practices to allow research to be readily discoverable and amenable to incorporation into simulations. We argue that there are aligned incentives in widespread adoption of practices which will both advance the needs of integrative simulation efforts as well as other contemporary trends in the biological sciences, ranging from open science and data sharing to improving reproducibility.

## Introduction

Theoretical research in the biological sciences has experienced accelerated growth over the course of the 20th and 21st centuries. From population genetics,[1] to macromolecular polymer dynamics,[2] to theoretical neuroscience,[3] mathematical modeling and fundamental theory have slowly crept into many areas of biological research. In other areas of science and engineering, the growth of mathematical techniques has paralleled the rise of computers as playing a fundamental role in the research process. However, simulations have had a controversial and varied history in the biological sciences. In particular, the diversity of efforts aimed at simulating complex biological systems, whether whole cells, simple organisms, tissues, or organ systems, have received nearly uniformly muted responses by the wider community of biological researchers.

We term these efforts "integrative biological simulations" because they integrate diverse, process-specific models into larger, composite models often operating at substantially different scales. Examples include WholeCell, a computational model of the bacterial parasite *Mycoplasma genitalium* integrating a broad range of dynamic, intracellular models such as transcription regulation, ribosome assembly, and cytokinesis[4,5] Open-Worm, an international, collaborative open-science project working towards a realistic, biophysical simulation of both the nervous system and body movement of *C. elegans*[6,7]

BlueBrain, an effort to build a detailed simulation of a rat cortical microcolumn[8] NeuroKernel, an analogous project to OpenWorm for *Drosophila melanogaster*[9] VirtualRat, a research program aimed at modeling the cardiovascular system of the rat and the related multi-decade efforts of physiologist Denis Noble and colleagues to mathematically model and simulate the human cardiovascular system[10,11]; ComputablePlant, an effort to develop a quantitative, cellular description of development in *Arabidopsis thaliana*[12]; and finally, Virtual Cell, a general computational framework for cell biological modeling.[13]

The list given above is simply a subset of the many diverse research efforts conducted over a several decade period. The fact that such projects continue to be attempted, and in such an incredible array of biological systems, suggests that there is a shared vision that continues to inspire researchers. Yet many scientists question their value, and simulations have not yet achieved the success in biology that they have in other areas of science, most notably in physics.[14–16] What should we make of this state of affairs? On the one hand, the vision is clear. Powerful computers should allow us to tame biological complexity. On the other, one wonders if this is enough. Is it possible that biological systems are sufficiently complex that attempts to incorporate more and more detail into massive computational models are fundamentally misguided? Or is it simply that despite several decades of effort, we are still in the very early stages of building an infrastructure that will have a deep and lasting impact on biological thought?

Our position on this matter is very much in the latter camp. We believe that integrative biological simulations are in their infancy and that their real successes lie in the future. In this commentary, we defend this position by engaging both philosophical and practical issues related to the simulation of complex biological systems. As physics has had the longest standing tradition of mathematical theories in the natural sciences, we begin by examining several sub-disciplines (atomic physics, particle physics, and astrophysics) in which simulations have played an integral role in the research process. We evaluate these subjects by the degree to which theory, simulation, and experiment interact with one another and the role that cost of research and speed of experimental feedback play in enabling integration. Our aim is to provide analogies from long-standing research traditions to help paint a picture of the role simulations may ultimately come to play in the biological sciences.

Next, we examine fundamental philosophical questions that consistently arise in discussions about simulations, such as the ontological status of biological theories and their relationship to the prospects of integrative simulation. We focus the discussion around the concepts of phenomenological and reductive modeling and argue that the success of biological simulations should be evaluated on a short and long-term basis. In the short term, their primary purpose is data and model integration and the reproduction of pre-existing knowledge. In the long term, we argue that simulations will come to be organically integrated into biological thinking analogous to some areas of physics.

Finally, we examine the critical role that data and model curation play in modern biology. In the same way that raw computational power has been a limitation and key enabling factor for computational methods in other areas of science, we argue that a culture of data and model curation is a limiting reagent in the biological sciences and is inadequately developed to allow for the flourishing of integrative simulations. Moreover, we propose a principle of *incentive structure alignment*, that many of the constituent practices required to execute integrative simulations are also relevant to other important contemporary trends in the biological sciences. When these practices are more widely distributed across the entire community, not only will a linchpin element of integrative simulations be enabled, but these efforts will also be less ambitious and instead, perceived to be a more natural outgrowth of collective research output.

We state at the outset that it is not our aim in this commentary to address every criticism that has been raised on this topic. The issues involved are simply too vast for a single article to cover. In particular, we do not enter into technical discussions of specific theoretical challenges faced by model building in biology. Likewise, our examination of philosophical issues related to simulations is not intended to be a contribution to fundamental thought on the topic, but instead, a compact summary of important themes that we have seen consistently arise in our discussions with others. Our aim is to take a bird's eye view of several key issues, both philosophical and practical, that we see as crucial components of a coherent perspective on the challenges and potential of integrative biological simulation.

## Research cost and speed of feedback—analogies from physics

Simulations can play an integral and organic role in the future of theoretical and experimental research in biology. To motivate this perspective, we describe several areas of physics, namely, atomic, molecular and optical physics (AMO), high-energy particle physics, and astrophysics / cosmology, in which simulations have been a fundamental part of day-to-day research for many years and decades. We compare the different roles that simulations play in these areas and discuss the relationship between simulations and the cost of conducting experiments. We argue that existing areas of physics can provide some intuition for how simulations might become a standard tool of biological research accessible to the whole community and which can shape biological thinking. Specifically, we point to examples of research where despite highly precise theories amenable to simulation, prohibitively expensive experiments can limit the rate of feedback to theoreticians, ultimately preventing the development of a research culture in which theory, simulation, and experiment co-exist in an integrated fashion. Conversely, precise theories, low experimental cost, and rapid feedback can allow for a more integrated culture in which simulations play a crucial supporting role in enabling theory and guiding future experiments.

Atomic, molecular, and optical physics (AMO physics), high-energy particle physics, and astrophysics / cosmology are unique disciplines in science. Few subjects are built upon theories-in these instances, quantum field theory and general relativity-that have been so thoroughly validated that many quantities are known to unfathomable degrees of precision. In recent years, scientists and the public alike have been enthralled by the phenomenal success of these theories in predicting the existence of the elusive Higgs Boson, a prediction made in the 1960's, and more recently, gravitational waves, a prediction first made by Albert Einstein himself a century ago in 1916.[17–19]

Although much of the history of research in quantum mechanics and general relativity took place before the advent of computers, in recent years, simulations have come to play an integral role in guiding both theoretical and experimental research. In astrophysics, for example, simulations have been used to model the stellar systems, namely black holes colliding, that emitted the gravitational waves detected by the LIGO collaboration.[20–22] In particle physics, simulations are used to model experiments conducted at particle colliders. For instance, in the years leading up to the first runs of the Large Hadron Collider (LHC) at CERN, a series of "LHC Olympics" were conducted. In these "events," simulated data of possible experimental outcomes were given to theorists in order to familiarize them with the practical issues that they would face in comparing the actual experimental results to potential theories once runs of the LHC began.[23,24] In both cases, astrophysics / cosmology and high-energy particle physics, the fundamental theories are of such enormous precision that extremely costly and delicate experiments can be conducted to push the limits of our understanding. And simulations of the underlying equations that characterize these theories have played a crucial role in many aspects of the research process.

We wanted to compare and contrast these subjects with AMO physics, a subject which is also founded upon quantum field theory, but which unlike high-energy particle physics, is concerned with the interaction of atoms and fields at energy scales that can be achieved with table-top experiments.[25,26] While the cost of particle colliders and gravitational wave detectors can range in the billions of dollars and involve thousands of scientists and engineers, experiments in AMO physics typically cost on the order of hundreds of thousands of dollars and laboratories of a dozen or so graduate students, post-docs, and research scientists. Still, AMO physics enjoys the phenomenal correspondence between theory and experiment shared by the set of disciplines built upon general relativity and quantum field theory. One well-known application of AMO physics is in the field of precision measurement and atomic clocks. Consider that the most precise atomic clocks, for example, are able to keep time with a stability of 1 part in $10^{18}$.[27,28]

However, while AMO physics, high-energy particle physics, and astrophysics / cosmology all share theoretical foundations and astonishingly strong theoretical / experimental correspondence, the low cost of experiments in AMO physics means that theory, experiment, and simulations are much more tightly integrated than in the other subjects. As an example, there are many purely theoretical papers which are not directly tested, simply because researchers can be confident on the basis of their basic understanding and simulations that a more ambitious experiment should be done instead which will implicitly test a range of smaller theories. In addition, it is not uncommon to find experimental laboratories which are also staffed with full-time theorists. Conversely, there are researchers exclusively trained as theorists who have started experimental research groups, typically with the assistance of full-time research scientists who were trained in experimental laboratories. In contrast, due to the substantial infrastructure required to conduct experiments, high-energy particle physics and astrophysics / cosmology operate on much longer time frames. Primarily for this reason, there is little direct overlap between the day-to-day research of theorists and experimentalists, and these research groups largely interact via the results of major experiments, rather than within a single research group.

We hope that the descriptions given above of the relationship between theory, simulation, and experiment in several areas of physics provide a perspective on the diverse ways in which these three approaches can be integrated in different disciplines. While simulations in biology have some commonality with simulations in other fields, such as the areas of physics given above, there are several key differences. Most significantly, biological systems, whether individual cells or the simplest multicellular organisms, are significantly more complex than individual atoms or large stellar systems.[29–32] Therefore, precise simulation requires integrating models that operate at very different scales involving many different parameters. Furthermore, the underlying theories of each of these models are unlikely to achieve the level of correspondence with experiments that we see in the most successful areas of fundamental physics. Although simulations in the areas of physics discussed above can also involve models at different scales, there is often precise theoretical understanding of the relationship between the constituent models, ultimately reducing the complexity of an integrated simulation. Nonetheless, there is an important role for simulations in cellular and organismic biology that is distinct from the areas of physics we described above—namely, data and model integration as well as outcome prediction. While simulations in a subject such as AMO physics may involve a single model specified by a handful of parameters, the whole-cell *Mycoplasma genitalium* simulation of Karr et al. required 28 models corresponding to some 1900 experimentally determined parameters.[4] Again, this is a manifestation of the level of complexity of biological systems when compared to single atoms or large stellar systems. This type of simulation, one whose role is to integrate diverse models and data sources, is distinct from those that are run in the branches of physics that we have described. The implications for future experimental research are significant. By combining the theoretical and experimental work of an entire community, integrative simulations will allow for significantly deeper interaction between these otherwise disparate sub-communities. Indeed, we should imagine a future where single research groups are composed of experimentalists, theorists, and computational model builders working side by side. Although theories in biology are unlikely to achieve the same success as theories in physics, due to the relatively low cost of experiments and speed of feedback, we can imagine that cellular and organismic biology has the potential for integration of these diverse research approaches along the lines of atomic physics. The ultimate consequence of this integration is that simulations will allow for hypothesis generation and selection, motivate novel experiments, and be organically integrated into biological thinking itself, thereby providing an ever-evolving representation of the collective state of knowledge in each field.

**Summary**: *Although physics is often cited as a subject where simulations have been successful, a more granular examination reveals that the role of simulations can vary substantially between sub-disciplines. The level of integration between theory, experiment, and simulation is strongly influenced by the speed of feedback from experiments, which is in turn influenced by the cost and complexity of those experiments. In imagining the role that simulations may come to play in biology, it is worth considering which sub-disciplines may serve as cultural analogies. We have singled out AMO physics as providing some useful intuition for reasoning about the future of integrative simulation in biological research.*

## Pragmatic philosophy of models and simulations

As in physics, discussions of the role of models in biological research often touch upon fundamental philosophical questions. What is the role of theory in biology, and when it succeeds, what status should we ascribe to mathematical models? Some have argued that it is fundamentally misguided for biology to aspire to achieve the same success with predictive models as physics and engineering. For example, in the essay "Biology is More Theoretical than Physics,"[33] Gunawardena gives a beautiful analysis of the origin of the Michaelis-Menten equation and observes that while Michaelis was the first to recognize the importance of pH for regulating enzymatic activity, as well as to develop techniques for buffering reactions, the Michaelis-Menten model does not have any pH dependence. While their model is an accurate description of a set of features in highly controlled experiments, it is not a realistic description of the many complexities of the experimental setup or the many crucial physical determinants of enzyme reactions such as pH.

Gunawardena draws a bold conclusion from this analysis, namely that *models are not descriptions of reality; they are descriptions of our assumptions of reality*. This is not a vacuous philosophical stance, but one with practical implications. He argues that this viewpoint should direct us away from attempting to engineer computational models with more and more details, and instead focus our efforts on models which are determined by the questions being asked and experimentally available data.

To address this philosophy and its relationship to biological simulations, we introduce several concepts from physics and philosophy of science, namely, the distinction between phenomenological and reductive models. A "phenomenological model" is one that attempts to model the observed, measured behavior of a system, without attention to underlying mechanisms. For instance, the Hodgkin-Huxley model of neuronal firing does not account for the complex cellular mechanisms underlying action potential generation—this complexity is reduced to a set of differential equations which capture the changes in axonal electrical activity. In other words, as the name implies, a phenomenological model simply attempts to "capture the phenomenon." Likewise, as the analysis of enzyme kinetics referenced above demonstrates, the Michaelis-Menten equation is a phenomenological model, one which describes the relationship between several observed variables, but leaves out an enormous amount of detail of the physical environment in which the enzymatic reaction takes place.

On the other hand, a reductive model is one which both accurately reproduces the observed behavior *and* the underlying mechanisms. Suppose we had a more detailed model which reproduced the behavior of the Hodgkin-Huxley model, but which was significantly more complex by virtue of being more complete. Would we use it instead? We may very well choose not to if it does not add any value. In other words, the practical distinction between reductive and phenomenological models amounts to a principle of parsimony. We want to incorporate only as much detail as is necessary to gain insight into the behaviors that we are interested in. This philosophy was most forcefully put by the eminent physicists Nigel Goldenfeld and Leo Kadanoff when they said "don't model bulldozers with quarks."[34] In other words, even though we know quarks to be fundamental constituents of the composition of matter, they exist at a level of abstraction well beneath what is necessary to model macroscopic objects. There is no need to incorporate this additional level of detail into our models. Although Goldenfeld and Kadanoff's dictum is stated quite differently than that of Gunawardena—they make no mention of the ontological status of models, for instance—it is often argued at dinner table discussions that the implication of this philosophy for computational modeling in the biological sciences is similar.

We very much acknowledge that if there is a cautionary tale to keep in mind, it is that the allure of powerful computers might seduce us into being less parsimonious than we ought to be in our modeling efforts. However, we are also cautious of allowing philosophical positions on the status of models to unduly inform one's position on the value of integrative simulations. Discussions concerning the relationship between symbolic representations of natural phenomena and computation go back at least as far as the European scientific revolution.[35,36] During this time period, Gottfried Leibniz and a number of his contemporaries pursued the development of a "universal calculus," a symbolic language which would represent all knowledge, thereby allowing computation to be a fundamental element of systematic thinking in areas as diverse as law, theology, and physics.[37] As science has progressed, the parallel philosophical inquiries have developed equally in their richness and diversity. In the 20th century, the dramatic successes of theoretical physics created the context for Wigner's seminal essay "On the unreasonable effectiveness of mathematics in the natural sciences."[38] And more recently, the tidal wave of advances of in machine learning inspired Halevy, Norvig, and Pereira's essay "The unreasonable effectiveness of data."[39] The latter case is particularly interesting in the present discussion. Models in machine learning might be considered the ultimate phenomenological models. Whereas the mathematical models described above consist of compact equations whose variables are recognizable quantities, machine learning models are often completely opaque and are effectively treated as inscrutable, black box components of larger software systems. Therefore, for those who see the fundamental issue with integrative simulation as relating to a tension between phenomenological and reductive modeling, it is worth pointing out that when taken to its extreme, phenomenological modeling gives rise to a class of techniques, which while practically useful in many ways, are not particularly appealing from a more general principle of parsimony. We mention these examples to emphasize that our philosophical understanding of mathematical models is continually evolving. In examining the breadth, depth, and historical trajectory of these philosophical questions, it seems that we are unlikely to arrive at decisive answers without fundamental theoretical advances on many fronts. Therefore, we adopt a pragmatic stance and advocate continuing philosophical discussions alongside more concrete objectives where possible.

We maintain a similarly pragmatic attitude towards Goldenfeld and Kadanoff's memorable quip concerning bulldozers and quarks. From a philosophical perspective, there are clearly deep insights which underlie this sentiment, and which almost certainly relate to Wigner's question. However, it is not clear that we can obtain any practical guidance from this perspective. For one thing, it would be equally misguided to model bulldozers with atoms, molecules, or polymers—one need not jump all the way down to the level of the quark. Yet, in progressing from quarks, to atoms, to molecules, and to polymers, we must move up many orders of magnitude in length scale. Our understanding of when to stop almost always comes from intuition and experience with a particular system rather than from fundamental principles. Moreover, it is a woefully inaccurate analogy to characterize integrative biological simulations as modeling bulldozers with quarks. Perhaps a better analogy would be to model bulldozers with pistons, shafts, chassis, and wheels. Surely such a level of detail is entirely appropriate to understand a bulldozer.

What then is the purpose of integrative simulations in the biology? Our viewpoint is that the answer depends on the time frame: there is a near-term and a far-term purpose. We believe that the purpose of simulations in the near-term is for *model and data integration*. We do not believe that integrative simulations introduce a conflict between phenomenological and reductive models. Phenomenological modeling has a rich history in theoretical biology and substantial work in building large-scale simulations involves curation, assembly, and infrastructure development for the management and integration of these modular sub-components. Indeed, the idea of modular modelling as a basis for an ecology of models and experiments is not a new one.[40-42] However, these efforts have always been limited and the fundamental principles have yet to permeate across the entirety of the biological sciences. When those processes are more systematically incorporated into the research process of an entire community, integrative simulations themselves will be less ambitious efforts. Their *ultimate* purpose, therefore, when the platforms have sufficiently matured, is to be organically integrated into biological thinking in the way that simulations have been woven into the fabric of research in AMO physics. They are *not* an excuse to *not* think deeply about which details of cellular function to incorporate and which to ignore. *Nor are they substitutes for experiments*. Rather, they should allow us to integrate the work of an entire community of researchers, theorists and experimentalists alike, into a collective structure, thereby revealing behavior that no human could reasonably extract through thought and experimentation alone.

Critics of large-scale simulations have argued that no novel or decisive insights into biological function have emerged from these projects.[14,15] However, if we were to survey researchers in multiple areas of theoretical physics we would likely hear a similar claim. No fundamental advances in atomic physics and quantum optics, for instance, have resulted from simulations alone, yet they have played an important supporting role in both theoretical and experimental research. Some have also argued that highly specific, phenomenological models in biology, where simulation played no role at all, have at times predicted the existence of entities or processes years or decades in advance of their experimental confirmation. As Gunawardena argues,[33] Michaelis and Menten's original derivation of their eponymous equation presumed the existence of a hypothetical entity, the enzyme-substrate complex, whose existence was not confirmed for 30 years. However, the whole-cell simulation of Karr et al. also makes a novel prediction, that genomic replication of *Mycoplasma* is rate-limited by cellular free deoxyribotrinucleotide phosphate (dNTP) at the onset of replication rather than at the beginning of the cell-cycle.[4] Likewise, the cortical micro-column simulation of BlueBrain makes a novel prediction concerning the transition from synchronous to asynchronous firing and calcium ion concentration.[8] Although these hypotheses have yet to be confirmed experimentally, we do not yet know what role these theorized processes and observed behaviors, and perhaps many others like them, will play in the thought processes of cell biologists and neuroscientists when such simulations are more widespread. Not only are they novel predictions, they are the result of massively novel engineering undertakings. And as we discuss in more detail below, efforts to develop integrative biological simulations are far more labor intensive than they need to be, as important components of these research programs could be distributed across an entire community and would benefit all researchers, regardless of their involvement with computational modeling. Therefore, if there is an unrealized potential for simulations in the biological sciences, it is this: a bridge between theory and experiment that will play a supporting role in guiding the research process and advancing biological thought.

**Summary**: *Both in the research literature and in our discussions with scientists, a common theme we have observed is that individuals' opinions on the potential of integrative biological simulation are often informed by philosophical positions concerning the status of theories and models in science. Yet outside of academic philosophy, these viewpoints are often not subject to the same level of intellectual scrutiny and debate as other scientific issues. Considering the diversity of positions on these topics and their rich intellectual history, we are cautious of either advocating or dismissing technical research agendas on philosophical grounds. Rather, we endorse a pragmatic stance of continuing concrete technical work while simultaneously acknowledging the importance of the underlying philosophy and openly stating and discussing one's philosophical beliefs.*

## The importance of data and model curation

In this section, we discuss key concepts and community practices for facilitating collaboration between the rich diversity of research groups in the biological sciences. We place these practices under the broad heading of "data and model curation." We use the following operational definition of curation: *the process of manual oversight in the collection, annotation, and dissemination of information sources, whether data, models, or other research output, to ensure quality, reliability, and usability*. These suggestions are quite general and will have a positive impact for all areas of inquiry, and not just research which is of relevance to biological simulation. In other words, there is substantial *incentive structure alignment* between practices relevant for simulation as well as other contemporary trends in biological research. Indeed, many of the practices we describe here have been adopted in the genetics community due to the explosion of data from low-cost gene sequencing.[43,44]

## Publishing and citing standalone datasets

The availability of experimental datasets is a fundamental component of model building and validation. In recent years, important developments relevant to the sharing of datasets have come from the software industry. The growth of the startup ecosystem in Silicon Valley and elsewhere has given rise to many small companies working towards building tools for scientific research and collaborative workflows. Many of these services are inspired by tools that originated in the software industry, such as the use of version control for collaborative documented editing, or making research code publically available through code sharing websites such as GitHub.[45-47] Of relevance to integrative simulations, and the biological sciences more broadly, is the ability to publish standalone datasets. In addition to being published, it is now possible for DOI's to be assigned to datasets themselves, analogous to research articles. Figshare, Zenodo, and the Wolfram Data Repository are a few of the services which currently exist for hosting standalone datasets which are assigned DOIs. The publication of standalone datasets not only allows for more rapid discovery of novel insights by the broader research community, but crucially, it also means that the original research group which generated the data can be appropriately credited for the fruits of their labor.

## Facilitating discovery of research results and datasets via keywords from widely used taxonomies such as Gene Ontology

The ability to publish standalone datasets gives rise to several issues with regards to discoverability. Because standalone datasets do not need to be accompanied by corresponding research articles, it may be that datasets of widespread interest are not easy for other researchers to discover. One strategy to address this issue is the development of search engines based on curated datasets. Successful examples include the gene expression datasets curated under the NCBI search engine GEO DataSets or GEO Profiles. These engines allow dataset search by keywords or gene plus keywords, respectively. Similarly, one proposal which we make to address this issue is for researchers to use keywords from Gene Ontology[48,49] as meta-data for their datasets when published in a standalone fashion. For example, researchers who upload standalone datasets to Wolfram Data Repository or Figshare, in addition to giving prose descriptions of the data, details about the conditions under which the data was collected and so on, might use Gene Ontology terms as the keywords for the dataset. This will allow theorists and modelers who are searching for data pertinent to their model building and model validation efforts to more easily discover datasets which might otherwise be difficult to find through ordinary text search. More broadly, we can ask a very general question that expands this perspective not just to data, but to any component of a research project. For a given paper, how can scientists quickly identify the reagents, tools, and other materials used in the production of the result in question? Ongoing work is aimed specifically at addressing this crucial need-The Resource Identification Initiative, for example, was launched with the aim of developing a set of database identifiers for any resource described in biological papers.[50]

## Curation and open science

The ability to readily share datasets creates many issues related to workflow and culture, some of which we have touched upon above. The set of practices required to make widely available data usable for others can be cast under the umbrella of data curation. As we described above, data curation refers to the process of manual oversight in the collection, annotation, and dissemination of data to ensure quality, reliability, and usability. The phrase typically has the connotation of workflows wherein iterative tasks are delegated to individuals after some amount of software-based automation for collecting and processing data.[51–54]

Data curation has become particularly visible in the context of open science projects. As an example, consider GalaxyZoo, a project out of the astrophysics community and the Sloan Digital Sky survey, to classify stellar images based on the shapes of the galaxies contained in the images. Unfortunately, this is a task where image processing techniques struggle (at least at the time of the project's beginning—advances have been made since then), and human oversight is required to achieve high levels of accuracy.[46] By creating a well-designed interface whereby volunteers, or "citizen scientists," can browse the site, click through images, and classify galaxies after a brief training period in which they are familiarized with the existing taxonomy, GalaxyZoo was able to classify over 50 million images in the first year, taking advantage of over 150,000 volunteers.

FoldIt is an example of a biological open science project that has generated a tremendous amount of data by taking advantage of large numbers of human volunteers. Employing a strategy known as "gamification," FoldIt is a puzzle video game which encodes complex challenges in protein folding. Top scoring solutions are then examined by experts to determine their validity and relevance to real-world problems in protein folding and crystallography.[55-57]

In the biological sciences, there has been increasing effort in the domain of "biocuration," and multiple conferences have been devoted specifically to bringing together the wider community of biologists, bioinformaticians, and software engineers involved in curation-related efforts. Examples can be found in the series of conferences held by the International Society for Biocuration.

The success of these projects and the need for human oversight emphasizes some key points about large datasets. It is often the case that the raw data is unusable without some amount of additional processing in which the data is annotated or classified according to some schema. In some cases, these tasks cannot be automated and require human intervention to be done accurately. When laboratories produce data, there may be implicit information about that data that is common knowledge in the laboratory, but which other researchers may not know about. Therefore, even significantly smaller datasets require a process of curation and quality control for datasets to be readily usable by others in the field.

The OpenWorm collaboration has made extensive use of data curation to build easily queried databases about *C. elegans* physiology.[6,58] These databases were assembled by mining the academic literature for information about neuron types, ion channels, and datasets with which to validate ion channel models. Although the amount of information in these efforts is significantly less than that of GalaxyZoo, one important point of commonality is that much of this work could not be automated by computer algorithms. Rather, systematic workflows have been put into place whereby individual researchers and a broader group of volunteers curate information about nematode physiology into formats that form the foundation for subsequent modeling efforts. As the culture of open science, sharing of datasets, and greater interaction between theorists, modelers, and experimentalists grows, it should become easier to manage such curation efforts. As an example, in OpenWorm, figures of I/V curves (current / voltage behavior for a given ion channel) are often used as the source of data with which to validate ion channel models. However, because these figures are simply images and not usable data, discretized data points must be extracted from the graphical image. This requires a workflow which encompasses:

1) Creating a list of all ion channels in *C. elegans* which are part of the computational model.
2) A corresponding list of publications containing the necessary I/V curves.
3) Explicit, step-by-step instructions for using tools to digitize the figures in these papers and uploading the resulting datasets to the OpenWorm data repositories.
4) Validation efforts to ensure the quality and integrity of the data produced from these workflows.

In the future, as it becomes easier and more commonplace to include datasets alongside publications themselves, curation efforts will become more streamlined and likely bypass the stage of needing to manually digitize static images. In other words, as data curation becomes integrated into the culture of the research process, there will be less of a need for larger, focused curation efforts simply because data that is produced by the community will start out in more usable and exchangeable formats.

## Facilitating discovery of research results and datasets and ensuring data quality through the journal review process

Much of what we have described above consist of practices that scientists themselves can employ to make their research output more discoverable and usable by others in the field. We also believe that journals can play a key role in this process. Journals and services that publish standalone datasets or alongside manuscripts can ensure that quality control of the data is part of the review process (the Wolfram Data Repository, for example, requires manual oversight and approval of submitted datasets). In addition to suggesting revisions to a manuscript, for example, a journal might request that authors submit datasets to public repositories where they are curated,[59] add additional keywords in the case that there is no accompanying manuscript, or likewise, if there are terms of relevance not contained in the manuscript text.

In the context of modeling, we can imagine that during the submission or review phase of a manuscript, that journals ask authors if the contents of their work—whether it is a purely theoretical model, a standalone dataset, or a manuscript with an accompanying dataset—would be of relevance to those working towards large-scale, integrative simulations. In these instances, it would seem reasonable to cast a wide net. Even if an author simply has a hunch that their work might be of relevance to those working on simulations, it should be sufficient for a journal to either add an additional keyword or have other means of discovering such papers through advanced search tools on their website.

The purpose in these cases is simply to narrow down the set of possible papers that will be part of larger curation efforts. Consider that the work of Karr et al. required reviewing 900 papers to accumulate the models necessary for their whole-cell modeling effort. Extending this paradigm to build a simulation of *E. coli*, for example[60] would expand the volume of literature substantially, perhaps by a factor of 10 or more.

## Model curation: Ensuring adequate annotation of theoretical models with pertinent details of associated experiments

Widespread availability of theoretical models of cellular components opens the door to "plug and play modeling," whereby researchers can build integrative simulations by taking advantage of pre-existing models in the literature. See, for example, existing model databases such as BioModels and the CellML Model Repository. Indeed, this is a key element of the workflow leading to the *Mycoplasma* simulation of Karr et al. described above. However, there is an important caveat to keep in mind when sharing models. To the extent that the details of a given model, such as fit parameters, are the result of specific experiments, it is essential that these details be part of the published model. Otherwise, we run the risk of researchers drawing erroneous conclusions from simulations which were cobbled together from models originating from wildly differing experimental conditions, and which therefore, have little connection to reality.[61] As we move towards community practices which encourage data and model sharing,[62–63] it is of utmost importance that all relevant parameters pertaining to experimental conditions and theoretical assumptions underlying the corresponding models are made completely explicit and transparent. The standards being developed by the Resource Identification Initiative, for example, are particularly relevant to addressing this issue.[50]

## Enabling model discovery and interoperability

Thus far, we have focused on best practices for the sharing of datasets, ensuring data quality, and facilitating discoverability. We close with a brief discussion of issues related to standardization and discoverability of biophysical models. As we have described above, the modeling world is highly heterogeneous, and the nature of the mathematical tools used to describe systems in neurobiology, biochemistry, and developmental biology can vary significantly.

One consequence of this cultural diversity has been the emergence of multiple, redundant standards for different subdisciplines. COMBINE (Computational Modeling in Biology Network) is a multi-community initiative aimed at increasing the interoperability of the many overlapping standards for sharing computational models.[64] In a similar spirit, the Whole-Cell project has published guidelines and key concepts for facilitating whole-cell modeling and the sharing of systems biology models.[65,66] The development of standards, tools, and infrastructure for facilitating interoperability between existing ones will prove to be a major enabler of integrative modeling in the biological sciences.

We mention a simple proposal that is specifically of relevance to linking genes and datasets with models, namely, extending Gene Ontology to incorporate mathematical terms describing the corresponding models. By adding terms such as "ordinary differential equation," "partial differential equation,"

"non-linear partial differential equation," "linear programming," "dynamic metabolic simulation," and so on, genes can not only be described by the roles that their gene products play in a cell, but also by the mathematical methods used to model their behavior. It may be that the appropriate terms for incorporation into Gene Ontology should come from one of the pre-existing standards from the COMBINE initiative—this is a topic that merits further discussion by modelers and experimentalists alike.

It is worth mentioning that the number of terms that would be added to fully characterize mathematical models used in biophysics is likely quite small, perhaps less than a thousand terms. As a simple technique for estimating this value, we examined the number of index entries for two textbooks relevant to biophysical modeling, *Quantitative Biology* and *Stochastic Processes in Physics and Chemistry*.[32,67] Both books contained indices with between 500 and 1000 terms, and many of these terms would not be relevant for a mathematical addendum to Gene Ontology. Thus, we believe an estimate in the several hundred range is reasonable. Like the suggestions and projects described above, extending Gene Ontology to incorporate terms relevant to biophysical models will allow for greater discoverability of models, datasets, and manuscripts more broadly for all researchers in the biological sciences, and in particular, those working towards large-scale, integrative simulation.

**Summary**: *An integral and often overlooked component of biological simulations is data and model curation, which we define as the process of manual oversight in the collection, annotation, and dissemination of information sources, whether data, models, or other research output, to ensure quality, reliability, and usability. The utility of these practices is not restricted to biological simulation and their widespread adoption by the biological community would benefit many areas of research. Such a commonality between distinct research agendas is what we refer to as "incentive structure alignment." We view this notion as a potentially important principle for conflict-resolution in evaluating research for which sufficient scientific consensus has yet to emerge, as in the case of integrative biological simulation.*

## Conclusion

Although theory and mathematical modeling have a rich history in biology, the role of computational simulations has been more controversial and the fundamental philosophy less developed. The aim of this manuscript has been three-fold. We began by examining sub-disciplines of physics in which fundamental theories are extraordinarily precise and observed that the cost of experiments and speed of feedback are a crucial set of variables influencing the degree to which theory, simulation, and experiment are integrated. Next, we identified several philosophical arguments related to the epistemology and ontology of models in biology that consistently arise in discussions about the potential for integrative simulation. We have adopted a pragmatic stance of encouraging and participating in philosophical discussions in parallel with practical work. Finally, we observed that significant components of integrative simulation efforts are data and model curation, practices which will have widespread impact for many aspects of the modern research process independent of biological simulation. This is what we refer to as *incentive structure alignment*.

The dawn of sophisticated multi-scale models in cellular and organismic biology suggests an emerging interface for integrated theoretical, experimental, and computational research. Realizing this vision will require both an organic evolution of scientific philosophy as well as specific workflow and research practices. We hope this commentary has provided useful analogies from subjects which have had long-standing histories with both theory and simulation, practical philosophical discussion, and a list of community practices for furthering the culture of integrated and open research in the biological sciences.

## Disclosure of potential conflicts of interest

No potential conflicts of interest were disclosed.

## Acknowledgments

We would like to thank Stephen Larson, Charles Limouse, Piyush Kumar, Criss Hartzell, and several anonymous referees for insightful comments and critical reading of the manuscript.

## Funding

This work was supported by the Emory University School of Medicine Catalyst Grant to VF.

## ORCID

Gopal P. Sarma  http://orcid.org/0000-0002-9413-6202
Victor Faundez  http://orcid.org/0000-0002-2114-5271